\documentstyle[12pt,fleqn]{article}
\newcommand{\Frac}[2]{\frac{{\displaystyle #1}}{{\displaystyle #2}}}
\input epsf
\def\bea{\begin{eqnarray}}
\def\eea{\end{eqnarray}}

\def\fun#1#2{\lower3.6pt\vbox{\baselineskip0pt\lineskip.9pt
        \ialign{$\mathsurround=0pt#1\hfill##\hfil$\crcr#2\crcr\sim\crcr}}}
\def\re#1{{[\ref{#1}]}}
\def\reii#1#2{{[\ref{#1},\ref{#2}]}}
\def\reiii#1#2#3{{[\ref{#1},\ref{#2},\ref{#3}]}}
\def\reiiii#1#2#3#4{{[\ref{#1},\ref{#2},\ref{#3},\ref{#4}]}}
\def\eqr#1{{Eq.\ ({\ref{#1})}}}
\def\eqrr#1#2{{Eqs.\ ({\ref{#1},\ref{#2})}}}
\def\thesection{}
\begin{document}
\thispagestyle{empty}
\renewcommand{\thefootnote}{\fnsymbol{footnote}}

%%%%%%%%%%%%%%%%%%%%%%%%%%%%%%%%%%
%%%%%  INCLUDE THIS FOR FERMILAB PREPRINT   %%%%%%
%%%%%%%%%%%%%%%%%%%%%%%%%%%%%%%%%%
\font\ssqfont=cmssq8 scaled 2500
\font\sqfont=cmssq8 scaled 1100
\null\vspace*{-104pt}
\begin{center}
\parbox{8.0in}{
\epsfxsize=48pt \epsfbox{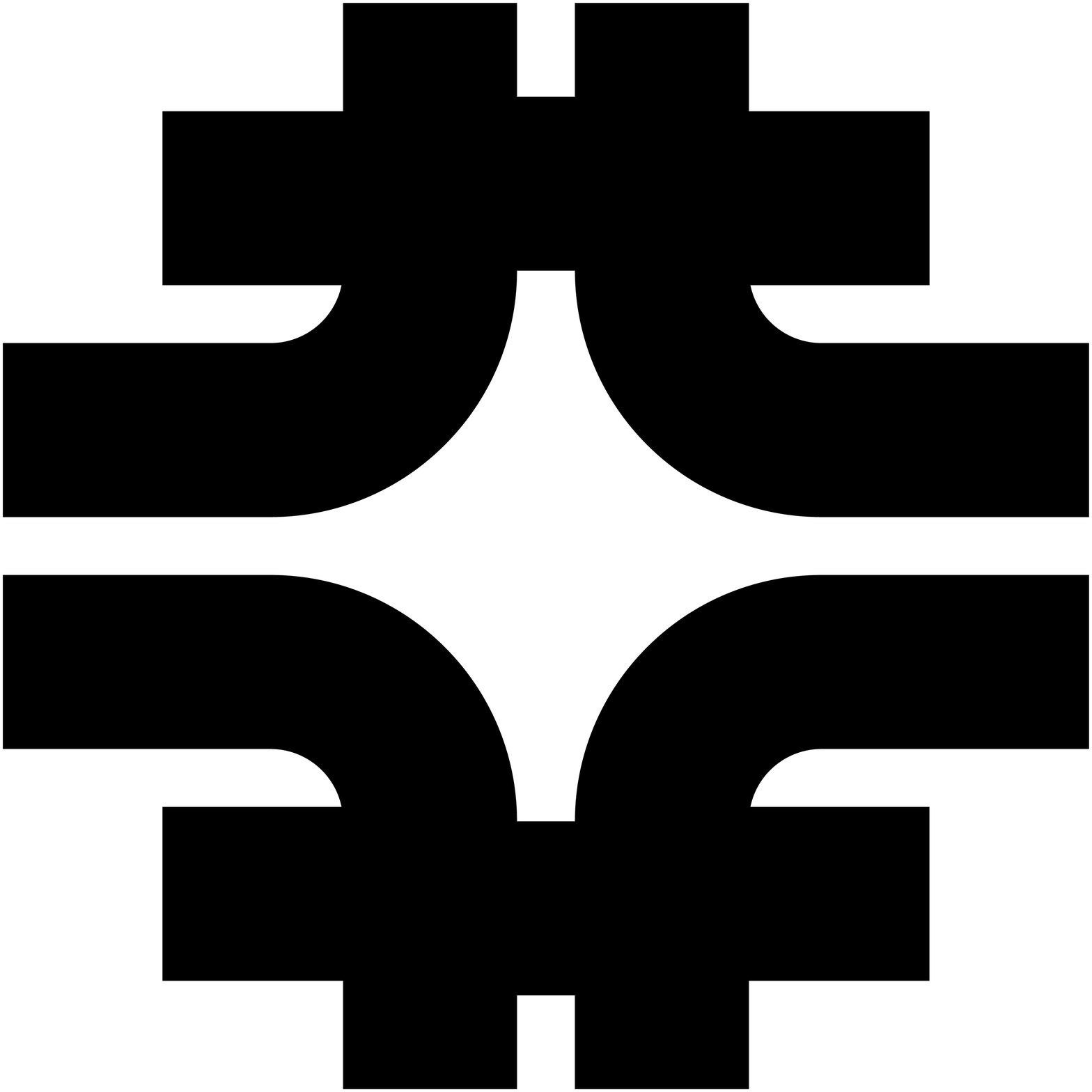}
\vspace*{-33pt}\hspace*{72pt}{\ssqfont Fermi National Accelerator Laboratory} 
}
\end{center}
\begin{flushright}
{\footnotesize
FERMILAB--PUB--97/037--A\\
astro-ph/9702166\\
\today}
\end{flushright}
\nopagebreak
\vspace*{8.5in}\centerline{
\epsfxsize=18pt \mbox{\epsfbox{Logo.eps} \ \  $^{\mbox
{\sqfont{Operated by Universities Research Association Inc.\ 
under contract with the United States Department of Energy}}}$} }
\vspace*{-8.75in}
%%%%%%%%%%%%%%%%%%%%%%%%%%%%%%%%%%
%%%%%%%%%%%%%%%%%%%%%%%%%%%%%%%%%%
%%%%%%%%%%%%%%%%%%%%%%%%%%%%%%%%%%
\baselineskip=24pt

\begin{center}
{\Large \bf Cosmic microwave background measurements \\
		can discriminate among inflation models}\\
\vspace{1.0cm}
\baselineskip=14pt

Scott Dodelson\footnote{Electronic mail: {\tt dodelson@hermes.fnal.gov}}\\
{\em NASA/Fermilab Astrophysics Center \\
Fermi National Accelerator Laboratory, Batavia, IL~~60510}\\
\vspace{0.4cm}
William H.\ Kinney\footnote{Electronic mail: {\tt kinneyw@fnal.gov}}\\
{\em NASA/Fermilab Astrophysics Center \\
Fermi National Accelerator Laboratory, Batavia, IL~~60510}\\
\vspace{0.4cm}
Edward W.\ Kolb\footnote{Electronic mail: {\tt rocky@rigoletto.fnal.gov}}\\
{\em NASA/Fermilab Astrophysics Center \\
Fermi National Accelerator Laboratory, Batavia, IL~~60510, and\\
Department of Astronomy and Astrophysics, Enrico Fermi Institute\\
The University of Chicago, Chicago, IL~~ 60637}\\
\end{center}

\baselineskip=24pt

\begin{quote}
\hspace*{2em} Quantum fluctuations during inflation may be responsible
for temperature anisotropies in the cosmic microwave background (CMB).
Observations of CMB anisotropies can be used to falsify many currently
popular models.  In this paper we discuss the prospectus for
observations of CMB anisotropies at the accuracy of planned satellite
missions to reject currently popular inflation models and to provide
some direction for model building.
\vspace*{12pt}

PACS number(s): 98.80.Cq, 98.80.Es,  98.70.Vc
\end{quote}

%\renewcommand{\thefootnote}{}
%\footnote{\raisebox{-36pt}{\ssqfont Fermi National Accelerator Laboratory}}
%\addtocounter{footnote}{-1}

\newpage
\baselineskip=24pt
\setcounter{page}{1}
\renewcommand{\thefootnote}{\arabic{footnote}}
\addtocounter{footnote}{-3}

%%%%%%%%%%%%%%%%%%%%%%%%%%%%%%%%%%%%%%%%%%%%
%%%%%%%%%%%%%%%  SECTION I    %%%%%%%%%%%%%%%%%%%%%%
%%%%%%%%%%%%%%%%%%%%%%%%%%%%%%%%%%%%%%%%%%%%
\thesection{\centerline{\large \bf I. INTRODUCTION}}
\setcounter{section}{1}
\setcounter{equation}{0}
\vspace{18pt}
%%%%%%%%%%%%%%%%%%%%%%%%%%%%%%%%%%%%%%%%%%%%
%%%%%%%%%%%%%%%%%%%%%%%%%%%%%%%%%%%%%%%%%%%%
%%%%%%%%%%%%%%%%%%%%%%%%%%%%%%%%%%%%%%%%%%%%

The field of observational cosmology has developed to the degree that
it is now possible to test models of the early universe.  The next few
years should see a dramatic increase in the variety and accuracy of
cosmological observations.  In this paper, we discuss how
measurements of the temperature anisotropies in the cosmic microwave
background (CMB) at the accuracy expected to result from two planned
satellite missions, the Microwave Anisotropy Probe (MAP) \re{MAP} and
PLANCK (formerly COBRAS/SAMBA) \re{CS}, will allow us to discriminate
among inflation models.

The basic idea of inflation is that there was an epoch early in the
history of the universe when potential, or vacuum, energy dominated
other forms of energy density such as matter or radiation. During the
vacuum-dominated era the scale factor grew quasi-exponentially while
the Hubble radius remained roughly constant.  Since in cosmic
expansion length scales increase with the scale factor, scales that
were once smaller than the Hubble radius grew during inflation to
become larger than the Hubble radius.  Once a length scale becomes
larger than the Hubble radius, any perturbation on that scale
becomes frozen.  Once inflation ends and the universe is radiation or
matter dominated, the Hubble radius increases faster than the scale
factor and the length scale reenters the Hubble radius with the
signature of events during inflation imprinted upon it.

By observing fluctuations in the CMB or fluctuations in the
distribution of matter, we can observe the signature of quantum
fluctuations during inflation.  Since different potentials lead to
different signatures, we can see which inflation models are consistent
with CMB fluctuations\footnote{We are assuming that inflation is
responsible for the anisotropies. There has recently been a large
amount of work \reiiii{CT}{ANDY}{HW}{TUROK}\ trying to understand how
easy it will be to distinguish anisotropies produced by inflation from
those produced by other mechanisms. We do not enter into this debate
here.}.
A problem with this effort \reii{KNOX}{KTURN}\ of trying to extract
information about the inflaton potential from the CMB is that the
anisotropies depend not only on the inflationary parameters, but also
on a variety of other unknown cosmological parameters
\reiii{COSCON}{JUNG}{ME}.  Among these are the baryon density
$\Omega_B$, the Hubble constant $H_0$, and the cosmological constant
$\Lambda$. Here, we fix the cosmological constant to zero. Allowing
$\Lambda$ and/or other parameters to vary would loosen the constraints
on inflationary models. On the other hand, we have not included
information that will be gained from measurements of CMB polarization
or from ongoing ground-based and balloon measurements of temperature
anisotropies.  So we expect our final constraints to be realistic
indicators of what we will know in ten years.

%%%%%%%%%%%%%%%%%%%%%%%%%%%%%%%%%%%%%%%%%%%%
%%%%%%%%%%%%%  SECTION II  %%%%%%%%%%%%%%%%%%%%%%%%
%%%%%%%%%%%%%%%%%%%%%%%%%%%%%%%%%%%%%%%%%%%%
\vspace{48pt}
\thesection{\centerline{\large \bf II. INFLATION DYNAMICS AND CMB 
FLUCTUATIONS}}
\setcounter{section}{2}
\setcounter{equation}{0}
\vspace{18pt}
%%%%%%%%%%%%%%%%%%%%%%%%%%%%%%%%%%%%%%%%%%%%
%%%%%%%%%%%%%%%%%%%%%%%%%%%%%%%%%%%%%%%%%%%%
%%%%%%%%%%%%%%%%%%%%%%%%%%%%%%%%%%%%%%%%%%%%

In this paper we consider only inflation models with ``normal''
gravity (i.e., general relativity) and a single scalar field (the {\it
inflaton}).  Although this might seem like a small region in the space
of possible inflation models, it does include some of the most studied
models, including scalar field models with polynomial
potentials ($\phi^n$), pseudo Nambu-Goldstone potentials (natural
inflation), exponential potentials (dilaton-like models), or
Coleman-Weinberg potentials (``new'' inflation).  In Section IV we
will describe the individual models we test.

In addition to the models we study, many other types of inflation
models can be studied by considering an equivalent one-field,
slow-roll model.  Two familiar examples are the Starobinski $R^2$
model and versions of extended inflation.  Both these models have
non-minimal gravitational sectors, with an $R^2$ term in the action of
the Starobinski model, and a Brans-Dicke coupling of $R$ to a scalar
field in extended inflation.  Although at first sight these do not
appear to be included in the class of models we study, after a
suitable conformal transformation both models can be expressed as
single-field, slow-roll inflation models. It would be interesting to
see if models with more than one inflaton field can be similarly
rewritten in terms of a single effective field. However, we do not
pursue this possibility here.

\vspace{24pt}
\thesection{\centerline{\bf A. Perturbation amplitudes and spectral 
indices from inflation}}
\setcounter{section}{2}
\setcounter{equation}{0}
\vspace{18pt}

Our goal is to start with a scalar field potential and calculate
the scalar and tensor perturbation amplitudes and spectral indices.
We make three basic approximations.  The first approximation is a
dual expansion of the metric about a Friedmann--Robertson--Walker
background, $g_{\mu\nu}({\bf x},t) = g^{\rm FRW}_{\mu\nu}(t) +
h_{\mu\nu}({\bf x},t)$, and an expansion of the inflaton field about a
homogeneous, isotropic background, $\phi({\bf x},t)= \phi_0(t) +
\delta\phi(\bf{ x},t)$.  Since we know that the density perturbations
are of order $10^{-5}$, this is presumably a very good approximation.

The metric perturbations produced by inflation can be described in
terms of two functions, which we call $A_S(k)$ and $A_T(k)$.  The
first function, $A_S(k)$, describes {\it scalar} metric perturbations.
These are the perturbations that couple to $T_{\mu\nu}$ and are
associated with structure formation.  The second function, $A_T(k)$,
describes {\it tensor} perturbations.  The tensor perturbations do not
couple to $T_{\mu\nu}$ and are not associated with structure
formation.  The tensor
perturbations can be visualized as gravity waves.  The distribution of
cosmic radiation depends on the full structure of the metric, so both
$A_S(k)$ and $A_T(k)$ contribute to CMB anisotropies.

The perturbation amplitudes $A_S(k)$ and $A_T(k)$ are the values the
quantities have when the wavenumber $k$ is equal to the Hubble
radius after inflation.  The scalar amplitude is related to the
density perturbation $(\delta\rho/\rho)_k$ and the power spectrum,
$P_S(k)\propto k^{-3}(\delta\rho/\rho)^2_k$, through a transfer
function $T(k)$ \re{transfer}.  We note that the normalization of
$A_S$ and $A_T$ is somewhat arbitrary, although the choice of
normalization will affect how we relate the parameters to directly
observable quantities; we follow the convention of Ref.\
\re{llkcba97}.  There $A_S(k)$ is normalized to be equal to the
density perturbation at Hubble radius crossing:
$A_S(k=aH)=(\delta\rho/\rho)_{k=aH} $.  The normalization of $A_T(k)$
was chosen such that to lowest order $A_T^2=\epsilon A_S^2$, where
$\epsilon$ is defined below.

The favored formalism for the calculation of perturbations uses the
Hamilton-Jacobi formulation of scalar field dynamics during inflation
\reiii{gs88}{m90}{sb90}, where the expansion rate, $H$, parameterized
by the value of the scalar field, $\phi$, is viewed as the fundamental
dynamical variable.
The most accurate calculations of the perturbation spectra are in
terms of $H$ and its derivatives.  The derivatives of $H$ can be
expressed in terms of dimensionless {\it slow-roll parameters}, the
first two of which are defined as\footnote{The definition of the
slow-roll parameters vary; we follow the conventions of Ref.\
\re{llkcba97}.}
\begin{equation}
\label{slowroll}
\epsilon (\phi) \equiv \frac{m_{{\rm Pl}}^2}{4\pi} \left( 
	\frac{H' (\phi) }{H(\phi)} \right)^2  ; \qquad
\eta (\phi) \equiv  \frac{m_{{\rm Pl}}^2}{4\pi} 
	\frac{H''(\phi)}{H(\phi)} .
\end{equation}
The second approximation we make involves the assumption that the
slow-roll parameters are small in comparison to unity. Note that
$\epsilon$ is a direct measure of the equation of state of the scalar
field matter, $p = - \rho \left(1 - 2\epsilon/3 \right)$, where $p$ is
the pressure and $\rho$ is the energy density. Since inflation can be
defined to be a period of accelerated expansion, where
\begin{equation}
\left(\frac{\ddot a}{a}\right) = H^2 \left(1 - \epsilon\right) > 0,
\end{equation}
the end of inflation can be expressed exactly as $\epsilon = 1$.

In the Hamiltonian-Jacobi formulation of the dynamics, the expansion
rate $H(\phi)$ is the fundamental cosmological parameter.  However, in
comparison with particle physics models, the inflaton potential
$V(\phi)$ is fundamental.  Thus, we have to express the slow-roll
parameters in terms of the inflaton potential.  This was done in Ref.\
\re{kv94}, with result
\begin{equation}\label{DEFEN}
\epsilon(\phi) = \frac{m^2_{Pl}}{16\pi}\left( 
\frac{V'(\phi)}{V(\phi)}\right)^2 \ ;  
\qquad \eta(\phi) =  - \frac{m^2_{Pl}}{16\pi}\left( 
\frac{V'(\phi)}{V(\phi)}\right)^2
	+ \frac{m^2_{Pl}}{8\pi}\left( \frac{V''(\phi)}{V(\phi)}\right) \ .
\end{equation}

The value of the scalar field can be used to specify a length scale
crossing the Hubble radius during inflation.  This is most easily
accomplished by considering the number of e-foldings of the scale
factor in the evolution from a value of $\phi$ until the end of
inflation: 
\begin{equation}
\label{ndef}
N(\phi,\phi_{\rm END}) \equiv \int_t^{t_{\rm END}}\! H(t) \, dt = \pm
 \frac{4\pi}{m^2_{Pl}} \int_\phi^{\phi_{\rm END}}\!
 \frac{H(\phi)}{H'(\phi)}\, d\phi \ , 
\end{equation} 
where the subscript `END' signifies that the quantity is to be
evaluated at the end of inflation. The choice of sign depends upon the
sign of $\dot{\phi}$, i.e., whether $|\phi_{\rm END}|$ is greater or
less than $|\phi|$.  It can be fixed by requiring the right-hand side
of the equation to be positive.

The comoving scale $k$ crosses the Hubble radius during inflation
$N(k)$ e-foldings from the end of inflation, where $N(k)$ is given by
\re{llkcba97}
\begin{equation}
\label{Ncross}
N(k) = 62 - \ln \frac{k}{a_0 H_0} - \ln \frac{10^{16} 
        {\rm GeV}}{V_k^{1/4}}
	+ \ln \frac{V_k^{1/4}} {V_e^{1/4}} - \frac{1}{3} \ln
	\frac{{V_e}^{1/4}}{\rho_{{\rm RH}}^{1/4}} \, .
\end{equation}
The subscript `0' indicates the present value of the quantity
and $\rho_{{\rm RH}}$ is the energy density after reheating.  For
instance, a length scale corresponding to $200h^{-1}$ Mpc (i.e.,
$k=2\pi/200h^{-1}$Mpc) roughly corresponds to $N(\phi,\phi_e) \simeq
50$.  Therefore the value of the inflaton field when a comoving scale
of $200h^{-1}$ Mpc crosses the Hubble radius during inflation is found
by finding $\phi_{\rm END}$ and solving \eqr{ndef} with $N(\phi,
\phi_{\rm END})=50$.

To lowest order in the slow-roll parameters, the scalar and tensor perturbation spectra are
\begin{equation}
\label{asat}
A_S(k)  \simeq  \frac{2}{5\sqrt{\pi}}  
\frac{1}{\sqrt{\epsilon(\phi)}}\frac{H(\phi)}{m_{Pl}}\ ;  \qquad
A_T(k)  \simeq  \frac{2}{5\sqrt{\pi}}  
\frac{H(\phi)}{m_{Pl}} \  .
\end{equation}
Note that the left hand side is expressed in terms of wavenumber $k$.
The relationship between $\phi$ and $k$ was discussed above.

 It is useful to describe the spectra in terms of spectral indices $n
\equiv d\ln A_S^2(k)/d\ln k$ and $n_T \equiv d\ln A_T^2(k) / d\ln
k$. Again to lowest order in the slow-roll parameters,
\begin{equation}
\label{indices}
n(k) - 1  \simeq  - 4 \epsilon(\phi) + 2 \eta(\phi)   \ , \qquad 
n_T(k)  \simeq  -2 \epsilon(\phi)  \ ,
\end{equation}
where once again it is necessary to use the relationship between $k$
and $\phi$.
A third approximation we make is that over the range of length
scales probed by CMB we can take the spectral indices as {\it
constant}.  In other words we assume that although the slow-roll
parameters change in inflation, they are roughly constant during the
epoch where scales of interest for the CMB cross the Hubble radius.
This implies that the scalar and tensor spectra can be written as
\begin{equation}
A_S(k)  =  A_S(k_0) \left( \frac{k}{k_0} \right)^{1-n} ; \qquad
A_T(k)  =  A_T(k_0) \left( \frac{k}{k_0} \right)^{n_T} \, ,
\end{equation}
where $n$ and $n_T$ are {\em constant} and $k_0$ is the wavenumber
corresponding to some length scale probed by CMB experiments.  This
allows the two {\it functions}, $A_S(k)$ and $A_T(k)$, to be
parameterized in terms of four {\it constants}, $\{A_S(k_0),\
A_T(k_0),\ n,\ n_T\}$.

If the perturbations arise from slow-roll inflation, then not all of
the four parameters are independent, but there is a relation, known as
the {\it consistency relation}, which reduces the number to three.  To
lowest order in slow-roll parameters, the consistency relation can be
found from Eqs.\ (\ref{asat}) and (\ref{indices}): $n_T=-2
A_T(k_0)^2/A_S(k_0)^2$.  So within the framework of the approximations
discussed above, the scalar and tensor perturbation spectra can be
characterized by three parameters, $\{A_S(k_0),\ A_T(k_0),\ n\}$.

\vspace{24pt} 
\thesection{\centerline{\bf B. Parameterization of the CMB perturbation 
spectrum}}
\setcounter{section}{2}
\vspace{18pt}

To calculate CMB spectra, one must solve the perturbed
Einstein-Boltzmann equations which describe how the different
components of the universe (photons, neutrinos, electrons, protons,
hydrogen, and dark matter) couple to each other and to gravity.  The
perturbation spectra produced by inflation are taken as initial
conditions for these equations. The final output is the full spectrum
of CMB perturbations. In Gaussian theories, such as inflation, these
are completely characterized by the two-point correlation function. If
the temperature pattern on the sky is expanded in spherical harmonics,
\begin{equation}
{\delta T(\theta,\phi) \over T_0} = \sum_{l=0}^\infty
\sum_{m=-l}^l a_{lm} Y_{lm}(\theta,\phi) 
\end{equation}
where $T_0=2.726$ is the average temperature of the CMB today, then
inflation predicts that each $a_{lm}$ will be Gaussian distributed
with mean zero and variance 
\begin{equation}
 C_l \equiv\langle \vert a_{lm} \vert^2 \rangle .  
\end{equation}
The $C_l$'s can be both measured experimentally and
predicted theoretically.

For a given set of inflationary parameters and cosmological
parameters, one can determine the full spectrum of $C_l$'s by solving
the Einstein-Boltzmann equations. Therefore, instead of specifying
thousands of $C_l$'s, it is more convenient to characterize a given
spectrum by the parameters which determine it. These are the three
parameters of the initial perturbation spectra, $\{A_S(k_0),\
A_T(k_0),\ n\}$ plus the unknown cosmological parameters, which we
take to be $\Omega_B$ and $H_0$. It has become conventional to
re-express the amplitudes $A_S(k_0)$ and $A_T(k_0)$ in terms of two
more physical quantities related to $C_2$. Specifically, we introduce
two parameters
\begin{equation}
Q_{rms-PS} \equiv T_0 \sqrt{ 5 C_2 \over 4\pi} \quad {\rm and} \quad 
r \equiv {C_2^{\rm tensor} \over C_2^{\rm scalar} }.
\end{equation}
Thus, any given set of $C_l$'s that we consider is a function of five
parameters, which we take to be $Q_{rms-PS},\ r,\ n,\ \Omega_B,$ and
$H_0$.

\vspace{24pt} 
\thesection{\centerline{\bf C. Connecting slow-roll
parameters and CMB parameters }} 
\setcounter{section}{2}
\vspace{18pt}

The natural parameters in ``model space'' are $H$, $\epsilon$, and
$\eta$, which correspond to the expansion rate during inflation, and
the first and second derivative of the expansion rate.  Since most
inflation models have an arbitrary adjustable parameter corresponding
to the normalization, information on the magnitude of $H$ is not as
valuable as information about the way $H$ changes. (Equivalently, no
theory predicts the value of $Q_{rms-PS}$.) So we find that
information about $\epsilon$ and $\eta$ gleaned from the harvest of
information expected from the next generation of CMB satellites will
be the best discriminant of inflation models. Here we relate
$\epsilon$ and $\eta$ to the observationally relevant parameters $n$
and $r$.

Equation \ref{indices}\ can be used to relate $n$ to $\epsilon$ and
$\eta$. The only ambiguity is the value of $\phi$ at which to evaluate
$\epsilon$ and $\eta$.  The most reasonable value of $\phi$ is the one
which corresponds to scales probed by the CMB. Thus, we define
$\phi_{\rm CMB}$ to be the value of $\phi$ associated with
$N(\phi_{\rm CMB},\phi_{\rm END}) = 50$.  (This is sometimes called
$\phi_{50}$ or $\phi_*$.)  By using \eqr{ndef}, $\phi_{\rm CMB}$ is
found from
\begin{equation}\label{DEFPCMB}
N(\phi_{\rm CMB},\phi_{\rm END}) = 50 = \sqrt{\frac{4\pi}{m^2_{Pl}} } \ 
\int_{\phi_{\rm CMB}}^{\phi_{\rm END}} \! \frac{1}{\sqrt{\epsilon(\phi')}} 
\ d\phi '\ .
\end{equation}
Then $n$ is given by
\begin{equation}
\label{scalarspectralindex}
n= 1 - 4\epsilon_{\rm CMB} + 2 \eta_{\rm CMB}
\end{equation}
where $\epsilon_{\rm CMB} \equiv \epsilon(\phi_{\rm CMB})$ and
similarly for $\eta$.

While the tensor to scalar ratio $r$ depends on $A_T(k_0)/A_S(k_0)$,
it also depends on $n$ somewhat, since $C_2$ coming from both tensors
and scalars is actually an integral over the primordial spectra.
Using fits to these integrals provided in Ref.\ \re{twl93}, it is
straightforward to show that, to lowest order in slow-roll,
\begin{equation}
\label{rofepsilon}
r = 13.7 \epsilon_{CMB} .
\end{equation}

We now have all the ingredients for a recipe to compare inflation
model predictions to CMB information.  The steps are:
\begin{enumerate}
\item From $V(\phi)$, calculate $\epsilon(\phi)$ and $\eta(\phi)$ 
using \eqr{DEFEN}.
\item Calculate $\phi_{\rm END}$ by $\epsilon(\phi_{\rm END}) = 1$.
\item Find $\phi_{\rm CMB}$ using \eqr{DEFPCMB}.
\item From $\epsilon_{\rm CMB}$ and $\eta_{\rm CMB}$ calculate $n$ from Eq.\
(\ref{scalarspectralindex}) and $r$ from Eq.\ (\ref{rofepsilon}), which can 
be compared directly to CMB temperature anisotropy data.
\end{enumerate}

%%%%%%%%%%%%%%%%%%%%%%%%%%%%%%%%%%%%%%%%%%%%
%%%%%%%%%%%%%  SECTION III  %%%%%%%%%%%%%%%%%%%%%%%%
%%%%%%%%%%%%%%%%%%%%%%%%%%%%%%%%%%%%%%%%%%%%
\vspace{48pt}
\thesection{\centerline{\large \bf III. SOME ONE-FIELD, SLOW-ROLL 
INFLATION MODELS}}
\setcounter{section}{3}
\setcounter{equation}{0}
\vspace{18pt}
%%%%%%%%%%%%%%%%%%%%%%%%%%%%%%%%%%%%%%%%%%%%
%%%%%%%%%%%%%%%%%%%%%%%%%%%%%%%%%%%%%%%%%%%%
%%%%%%%%%%%%%%%%%%%%%%%%%%%%%%%%%%%%%%%%%%%%

In this section, we look at several generic inflationary models. The
models we consider can be grouped into three general classes,
``large-field,'' ``small-field,'' and ``hybrid.''  Large-field models
are characterized by so-called {\em chaotic} initial conditions, in
which the inflaton field is displaced far from its minimum, typically
to values $\phi \sim m_{Pl}$, and rolls toward a minimum at the
origin. Examples A and E below are large-field models. Small-field
models are of the form that would be expected as a result of
spontaneous symmetry breaking, with a field initially near the origin
and rolling toward a minimum at $\left\langle \phi \right\rangle \neq
0$. In this case, inflation occurs when the field is small relative
to its expectation value, $\phi \ll \left\langle\phi\right\rangle$.
Examples B and C below are small-field models.

In order to avoid cumbersome notation we will assume that $\phi$ is
positive.  Clearly if the potential is an even function of $\phi$ then
the sign of $\phi$ is irrelevant, while if the potential is an odd
function of $\phi$, then $-V(-\phi)$ is equivalent to $V(\phi)$.

The large-field and small-field cases occupy very different regions in
the space of observable parameters, and can be formally distinguished
by the curvature of the potential in the region where inflation is
taking place: for the large field models, $V''\left(\phi\right) > 0$,
and for the small field models, $V''\left(\phi\right) < 0$. In
addition, we consider a fifth model (D) that sits on the boundary
between the small field and large field cases, which is the case of a
linear potential $V''\left(\phi\right) = 0$.

A third general class of models, occupying a distinct region of
parameter space, is ``hybrid'' inflation \reiii{al91}{al94}{cllsw94},
which is characterized by a field evolving toward a minimum of the
potential with a nonzero vacuum energy. Hybrid models generally
involve more than one scalar field, but can be treated during the
inflationary epoch as single-field inflation, with $\phi$ small and
$V''\left(\phi\right) > 0$. Hybrid inflation is the only class of
models which predicts a ``blue'' spectrum, $n > 1$. Case F below is a
generic hybrid model.

The idea is to be as general as possible, and we calculate the values
of observables as functions of parameters in the models avoiding
prejudices about the ``reasonableness'' of those parameters. For
example, it is possible that particular realizations of these cases in
more detailed contexts may require excessive fine-tuning or
implausibly large mass scales. However, a completely different model
may achieve the same behavior in a more natural way, and our goal is
inclusiveness. This results in particularly broad constraints in the
hybrid case. Hybrid inflation models as a class have enough adjustable
parameters that it is possible to generate observables covering broad
regions on the $\left(r,n\right)$ plane, and model-dependent physical
arguments must be invoked to limit the predictions. Nonetheless, even
with very weak assumptions, there is no overlap in parameter space
between hybrid inflation and the other cases considered.

\vspace{24pt}
\thesection{\centerline{\bf A. ``Large-field'' polynomial potentials: 
$\Lambda^4(\phi/\mu)^p$, $p>1$}}
\setcounter{section}{3}
\setcounter{equation}{0}
\vspace{18pt}

The simplest example of the type of inflation model we study is a
``large-field'' polynomial potential, $V(\phi) =
\Lambda^4(\phi/\mu)^p$ with $p>1$.  Here $\Lambda$ and $\mu$ are
parameters of mass dimension one; neither one enters in our
results.  This potential is often used in ``chaotic'' inflationary models
where some region of the universe starts with the scalar field
displaced from the minimum of the potential ($\phi=0$) by a large
amount, typically several times $m_{Pl}$, and evolves to the minimum.
In these models $\phi > \phi_{\rm END}$, so inflation occurs when the
scalar field is larger than its eventual minimum.

Following the steps outlined in Section II, we find:
\begin{enumerate}
\item The slow-roll parameters $\epsilon(\phi)$ and $\eta(\phi)$ are 
given by
\[
\epsilon(\phi) = \frac{p^2}{16\pi} \frac{m^2_{Pl}}{\phi^2}  ;  \quad
\eta(\phi) = \frac{p (p-2)}{16\pi} \frac{m^2_{Pl}}{\phi^2} \ .
\]
\item The end of inflation occurs when $\phi=\phi_{\rm END}$, given by
\[
\frac{\phi_{\rm END}^2}{m^2_{Pl}} =  \frac{p^2}{16\pi}  \ .
\]
\item The value of $\phi$ crossing the Hubble radius 50 e-folds from 
the end of inflation is 
\[
\frac{\phi_{\rm CMB}^2}{m^2_{Pl}} =  \frac{1}{16\pi} \ p (p+200) \ .
\]
\item The values of $\epsilon_{\rm CMB}$ and $\eta_{\rm CMB}$ are
\[
\epsilon_{\rm CMB} = \frac{p}{p + 200} \ ; \qquad \eta_{\rm CMB} = 
\frac{p-2}{p+200} \ .
\]
Using these values of $\epsilon_{\rm CMB}$ and $\eta_{\rm CMB}$, 
it is easily shown that
\[
n = 1 - \frac{2p+4}{p+200} \ ; \qquad 
r \simeq 13.7 \frac{p}{p+200}  \ .
\]
\end{enumerate}
Note that this is a minimalist model in the sense that inflation ends 
naturally, without the necessity of invoking another sector of the theory. 
The results are listed in Table 1.

\vspace{24pt}
\thesection{\centerline{\bf B. ``Small-field'' polynomial potentials: 
$\Lambda^4 [1 - (\phi/\mu)^p]$,  $\phi \ll  \mu \ll m_{Pl}$ and
$p > 2$} }
\setcounter{section}{3}
\setcounter{equation}{0}
\vspace{18pt}

The small-field polynomial describes what might result if the
potential arises from a phase transition associated with spontaneous
symmetry breaking.  In this scenario, the field is evolving away from
an unstable equilibrium at the origin toward a nonzero vacuum
expectation value, $\left\langle\phi\right\rangle \ne 0$. Near the
origin, the potential can be written as a Taylor expansion,
\begin{equation}
\label{smallfieldpotential}
V\left(\phi\right) = \Lambda^4 \left[1 - \left(\frac{\phi}{\mu}\right)^p + 
\cdots\right],
\end{equation}
where $p$ is the lowest non-vanishing derivative at the origin, and
$\mu \propto \left\langle\phi\right\rangle$. For instance, the
Coleman-Weinberg potential used in the original ``new'' inflation
models \reii{al82}{as82} is of this form with $n=4$. This ansatz is
quite general, applicable even to potentials which have a logarithmic
divergence in the leading derivative at the origin \re{km96}. In
keeping with the motivation for this model we will assume that $\mu
\ll m_{Pl}$, so we have the hierarchy of scales $\phi \ll \mu \ll
m_{Pl}$. The analysis was described in detail in \re{km96}; the
relevant results are given in Table 1 and illustrated in Fig.\ 1. (The
case $p = 2$ is special, and is discussed separately below.) Like the
polynomial large-field models, the parameters $r$ and $n$ are
independent of the fundamental mass scales in the potential,
\begin{equation}
r \simeq 0, \qquad
n = 1 - \frac{p - 1}{25 \left(p - 2\right)} \ .
\end{equation}
Unlike the large-field case, these models have the feature that 
$\epsilon_{\rm CMB}$, and hence $r$, is negligibly small.

\vspace{24pt}
\thesection{\centerline{\bf C. ``Small-field'' quadratic potentials:
$\Lambda^4[1-(\phi/\mu)^2]$, $\phi \ll \mu$}
\setcounter{section}{3}
\setcounter{equation}{0}
\vspace{18pt}

``Natural'' inflation models \re{NATURALINFLATION}, in which the
potential is usually assumed to have a cosine potential, can be
described by \eqr{smallfieldpotential} with $p=2$ near the origin
where inflation occurs.

Potentials dominated by a quadratic term have the property that the
small-field assumption $\phi \ll \mu$, while valid at the time when
observable parameters are generated, is not consistent all the way to
the end of inflation, since
\begin{equation}
\epsilon\left(\phi\right) = \frac{1}{4 \pi} \left(\frac{m_{Pl}}{\mu}\right)^2 
\frac{\left(\phi / \mu\right)^2}{\left[1 - \left(\phi / 
\mu\right)^2\right]^2} \ .
\end{equation}
Then $\phi_{\rm END} / \mu$ approaches unity for large $\mu$, and
higher order terms in the potential cannot be neglected. We adopt
the reasonable assumption that $\mu$ in some direct sense
parameterizes the expectation value of the field in the physical
vacuum, so that $\left(\phi_{\rm END} / \mu\right)$ is of order
unity. The precise value of $\phi_{\rm END}$ is not important, since
\begin{equation}
\label{phicmbquad}
\phi_{\rm CMB} = \phi_{\rm END} \exp\left[-\frac{25}{4 \pi} 
\left(\frac{m_{Pl}}{\mu}\right)^2\right]
\end{equation}
is exponentially small regardless, and the parameters $\epsilon_{\rm CMB}$ 
and 
$\eta_{\rm CMB}$ approach the small-field limits
\begin{equation}
\eta_{\rm CMB} = - \frac{1}{4 \pi} \left(\frac{m_{Pl}}{\mu}\right)^2,\qquad
\epsilon_{\rm CMB} = \left|\eta_{\rm CMB}\right| \exp\left[-100 
\left|\eta_{\rm CMB}\right|\right] \simeq 0.
\end{equation}
Note that since 
\begin{equation}
n = 1 + 2 \eta = 1 - \frac{1}{2 \pi} \left(\frac{m_{Pl}}{\mu}\right)^2,
\end{equation}
if $n > 0.9$ as suggested by the COBE measurements, then
$\mu \ll m_{Pl}$ is excluded. 
The scale-invariant limit is $\mu \rightarrow \infty$, or $\eta
\rightarrow 0$, but it is important to remember that the small-field
approximation breaks down in this limit, since $\phi_{\rm CMB}
\rightarrow \phi_{\rm END}$ in
\eqr{phicmbquad}.   

\vspace{24pt}
\thesection{\centerline{\bf D. Linear potentials: $\Lambda^4(\phi/\mu)$
and $\Lambda^4[1-\phi/\mu]$}}
\setcounter{section}{3}
\setcounter{equation}{0}
\vspace{18pt}

Linear potentials have the property that $\epsilon = -\eta
=m^2_{Pl}/16\pi\mu^2$ is {\em independent} of $\phi$.  Thus, if
inflation starts, i.e., if $\epsilon <1$, it will never end.  More
exactly, some other physics must enter to terminate the inflationary
phase.  So we assume that the linear potential is only valid when
scales of interest for the CMB are passing through the Hubble radius.
Thus the relevant values of $\epsilon$ and $\eta$ are those given
above. Like the quadratic potential, the scale-invariant limit is $\mu
\rightarrow \infty$.

\vspace{24pt}
\thesection{\centerline{\bf E. Exponential potentials: 
$\Lambda^4\exp\sqrt{16\pi\phi^2/pm^2_{Pl}}$\ , $p>0$}}
\setcounter{section}{3}
\setcounter{equation}{0}
\vspace{18pt}

Exponential potentials lead to an exponential form of the Hubble parameter, 
which in turn leads to a power-law time dependence of the scale
factor.  For potentials of the form $V(\phi) = \Lambda^4
\exp\sqrt{16\pi\phi^2/pm^2_{Pl}}$, the expansion rate is $H \propto
\exp\sqrt{4\pi\phi^2/pm^2_{Pl}}$ which gives $a \propto t^p$.  This
model is usually called power-law inflation, a term we will not use in
order to avoid confusion with models with power-law potentials.
Exponential potentials, while nonrenormalizable, arise quite naturally
as the effective low-energy description of degrees of freedom associated
with extra spatial dimensions in Kaluza--Klein models,  as well as dilatons 
and moduli fields in superstring theories.

This model has the useful property that both $\epsilon$ and $\eta$ are
constant and equal: $\epsilon =\eta = p^{-1}$.  Thus, as in the linear
potential case, some other physics must enter in order for inflation
to end.
With $\epsilon = \eta = p^{-1}$, we find $r = 13.7 p^{-1}$ and $n = 1 -
2p^{-1}$.  The result $n-1 \propto r$ is often incorrectly generalized
to all slow-roll models.

\vspace{24pt}
\thesection{\centerline{\bf F. Hybrid Inflation: 
$\Lambda^4 [1 + (\phi/\mu)^p]$,  $\phi < \mu$}
\setcounter{section}{3}
\setcounter{equation}{0}
\vspace{18pt}

The final class of models we consider is ``hybrid'' 
inflation \reiii{al91}{al94}{cllsw94}, in which the field rolls toward a 
minimum with a nonzero vacuum energy. We take a potential of the form
\begin{equation}
V\left(\phi\right) = \Lambda^4 \left[1 + 
\left(\frac{\phi}{\mu}\right)^p\right],
\end{equation}
with $p \geq 2$. The large-field limit of this potential is just the
case of chaotic inflation with a polynomial potential, model A. Hybrid
inflation is the limit of {\it small} field, $\phi < \mu$, where the
potential is dominated by the constant term, $V \simeq \Lambda^4 =
{\rm const.}$ In the absence of any other physics, the field rolls
toward the origin, coming to rest at $\phi = 0$ after an {\it
infinite} period of inflation. For inflation to end, another sector of
the theory must be invoked, generally a coupling to a second scalar
field $\psi$, so that $\phi_{\rm END}$ and $\phi_{\rm CMB}$ cannot be
fixed outside the context of a particular model. A generic
characteristic, however, is that $\phi_{\rm CMB} \gg \phi_{\rm END}$.
For generality, we will take $\left(\phi_{\rm CMB} / \mu\right)$ to be
less than unity; in many models it is often very much less than
unity. In hybrid inflation, the parameter $\eta_{\rm CMB}$ is
positive, and can be written in terms of $\epsilon_{\rm CMB}$
\begin{eqnarray}
\frac{\eta_{\rm CMB} }{\epsilon_{\rm CMB} }& = & 
\frac{2 \left( p - 1\right)}{p} \left(\frac{\phi_{\rm CMB}}{\mu}\right)^{-p} 
\left[1 + \frac{p - 2}{2 \left(p - 1\right)} 
	\left(\frac{\phi_{\rm CMB}}{\mu}\right)^p\right]  \cr
 & \longrightarrow & \left\{ \begin{array}{ll}  
\Frac{p-2}{p} & {\mbox {for\ }}\phi_{\rm CMB}/\mu \gg 1 \\  & \\
\Frac{2 \left( p - 1\right)}{p} 
\left( \Frac{\mu}{\phi_{\rm CMB}}\right)^p & 
		{\mbox {for\ }}    \phi_{\rm CMB}/\mu \ll 1 
\end{array} \right.   .
\end{eqnarray}
This first expression depends only on the assumption of slow-roll, not
on a small-field limit. In the large-field limit, $\phi_{\rm CMB}/\mu
\gg 1$, we recover the result for model A found above, $\eta_{\rm
CMB}/\epsilon_{\rm CMB} = (p-2) /p $.  In the small-field limit,
$\phi_{\rm CMB}/\mu \ll 1$ , we obtain the familiar result for hybrid
models, $n>1$.

This possibility of a
 ``blue'' scalar spectrum (here, blue implies $n>1$) is
the distinctive feature of hybrid models. Recalling that
 $n = 1-4\epsilon + 2 \eta$, we see that although hybrid models can in
principle result in a red spectrum (for $\eta < 2 \epsilon$), if
 $\eta > 2 \epsilon$, hybrid inflation predicts a blue spectrum.

The predictions for all of the models described here are summarized in 
Table\ 1.

\newpage

\begin{table}
\begin{center}
\begin{tabular}{c|c|c|c|c}
\hline \hline
 & & & &  \\
model & 
$\Frac{\phi^2_{\rm END}}{m^2_{Pl}}$ & 
$\Frac{\phi^2_{\rm CMB}}{m^2_{Pl}}$ & 
$\epsilon_{\rm CMB}$ & $\eta_{\rm CMB}$ \\
 & & & & \\
\hline \hline & & & & \\ 
A & 
$\Frac{p^2}{16\pi} $ &
$\Frac{p (p+200)}{16\pi}$ & 
$\Frac{p}{p + 200}$ &
$\Frac{p-2}{p+200}$ \\
 & & & & \\
B &
$\Frac{\mu^2}{m^2_{Pl}}
\left[ \Frac{\sqrt{16\pi}}{p}
\left( \Frac{\mu}{m_{Pl}}\right)\right]^{2/(p-1)}$ & 
$\Frac{\mu^2}{m^2_{Pl}}
\left[ \Frac{4\pi}{25p(p-2)}
\left( \Frac{\mu^2}{m_{Pl}^2}\right)\right]^{2/(p-2)}$ &
$\ll \left|\eta_{\rm CMB}\right|$ &
$-\Frac{p-1}{50(p-2)}$ \\
 & & & & \\
C &
$O\left(\Frac{\mu^2}{m_{Pl}^2}\right)$ & 
$\left(\Frac{\phi_{\rm END}^2}{m_{Pl}^2}\right) \exp\left[- \Frac{25}{2 \pi} 
\left(\Frac{m_{Pl}}{\mu}\right)^2\right]$ &
$\ll \left|\eta_{\rm CMB}\right|$ &
$-\Frac{m^2_{Pl}}{4\pi\mu^2}$ \\ 
 & & & & \\
D &
undetermined & 
undetermined &
$\Frac{m^2_{Pl}}{16\pi\mu^2}$ &
$-\Frac{m^2_{Pl}}{16\pi\mu^2}$ \\ 
 & & & & \\
E &
undetermined &
undetermined &
$p^{-1}$ &
$p^{-1}$ \\
 & & & & \\
F &
undetermined & 
undetermined &
$< \eta_{\rm CMB}$ &
$> 0$\\ 
 & & & & \\
\end{tabular}
\end{center}
Table 1: Lowest-order results for $\phi_{\rm END}$,  $\phi_{\rm CMB}$, 
$\epsilon_{\rm CMB}$, and $\eta_{\rm CMB}$ in some popular inflation models.
\end{table}

%%%%%%%%%%%%%%%%%%%%%%%%%%%%%%%%%%%%%%%%%%%%
%%%%%%%%%%%%%  SECTION IV %%%%%%%%%%%%%%%%%%%%%%%%
%%%%%%%%%%%%%%%%%%%%%%%%%%%%%%%%%%%%%%%%%%%%
\vspace{48pt}
\begin{center}
\thesection{\large \bf IV. EXTRACTING PERTURBATION 
SPECTRA INFORMATION  FROM CMB OBSERVATIONS}
\end{center}
\setcounter{section}{4}
\setcounter{equation}{0}
\vspace{18pt}
%%%%%%%%%%%%%%%%%%%%%%%%%%%%%%%%%%%%%%%%%%%%
%%%%%%%%%%%%%%%%%%%%%%%%%%%%%%%%%%%%%%%%%%%%
%%%%%%%%%%%%%%%%%%%%%%%%%%%%%%%%%%%%%%%%%%%%
\def\qrms{Q_{\rm rms-PS}}

Now that we know how to extract the observables
$n$ and $r$ from a given inflationary potential,
we turn to the question of how well experiments will be able
to measure these quantities. The general question of
parameter estimation from CMB experiments will likely occupy
cosmologists for a long time. However, without any simulations
at all, one can get a very good idea of how accurately
parameters will be determined by using a simple $\chi^2$
technique. A given experiment will measure each $C_l$
with an error given by $\Delta C_l$. The ``true'' set
of parameters
will be determined by minimizing 
\begin{equation} \label{CHI}
\chi^2\left(\left\{\lambda_i\right\}\right) \equiv \sum_{l=2}^\infty 
{ \left( C_l\left(\big\{\lambda_i\big\}\right) - C_l^{\rm measured} \right)^2
\over (\Delta C_l)^2 }.
\end{equation}
Here the set of parameters
$\left\{n,r,\qrms,\Omega_B,H_0\right\}$ which we are allowing to vary 
is denoted $\left\{\lambda_i\right\}$.

Of course, we cannot know in advance what $C_l^{\rm measured}$ will
turn out to be. But knowing what we expect for $\Delta C_l$, we can
get an estimate of how large the uncertainties in the parameters will
be.  To do this, we assume that the measured $C_l$'s will be very
close to the true $C_l$'s. Then, by minimizing the $\chi^2$, we will
accurately determine the parameters. Therefore, we can expand
\begin{equation}\label{EXCHI}
\chi^2\left(\left\{\lambda_i\right\}\right) \simeq
\chi^2\left(\left\{\lambda_i^{\rm true}\right\}\right)
+ {1\over 2} \left.
{\partial^2\chi^2 \over 
\partial\lambda_i\partial\lambda_j}\right|_{\lambda=\lambda^{\rm true}} 
\left(\lambda_i - \lambda_i^{\rm true}\right) 
\left(\lambda_j - \lambda_j^{\rm true}\right). 
\end{equation}
The second derivative matrix carries information about how quickly the
$\chi^2$ increases as the parameters move away from their true
values. Therefore, under some reasonable assumptions \re{PRESS}, the
uncertainties in the parameters are determined by this matrix. We will
be interested only in the parameters $n$ and $r$, so we want to project
these uncertainties onto the two-dimensional $n-r$ plane. (This is
equivalent to integrating out all the other variables.) It is a simple
exercise to show that these uncertainties are obtained by computing
the elements of the five-by-five second derivative matrix, inverting
it, and then picking out the two-by-two matrix corresponding to the
$n,r$ elements. The remaining two-by-two matrix defines the error
ellipses in the $n-r$ plane.

To complete this program, we need two more pieces of
information. First, the elements of the derivative matrix must be
evaluated at the true values of the parameters. We need to specify
what we are assuming for the true values. Here, we look separately at
two possible sets of values for the parameters. The first corresponds
to standard cold dark matter (sCDM).  
\begin{equation}
\Big\{\lambda^{sCDM}\Big\} =
\Big\{n,r,\qrms,\Omega_B,H_0\Big\} = \Big\{1,0,18\mu K,0.0125,50\Big\}
\end{equation} 
where $H_0$ is in units of km sec$^{-1}$ Mpc$^{-1}$.  The second
set corresponds to values of the parameters considered to be viable
upon consideration of large scale structure data \re{EQUI}.  
\begin{equation}
\Big\{\lambda^{LSS}\Big\} = \Big\{n,r,\qrms,\Omega_B,H_0\Big\} =
\Big\{0.9,0.7,18\mu K,0.02,50\Big\} \
\end{equation}
The $C_l$'s for these models
are shown in Fig.\ 1.  Since the anisotropies are considerably larger
in sCDM, the signal to noise in a given experiment will also be
larger. Therefore we expect tighter bounds in sCDM than in our second
model.

%%%%%%%%%%%%%%%%%%%%%%%%%%%%%%%%%%%%%%%%%%%%
%%%%%%%%%%%  FIGURE  1 %%%%%%%%%%%%%%%%%%%%%%%%%%%
%%%%%%%%%%%%%%%%%%%%%%%%%%%%%%%%%%%%%%%%%%%% 
\begin{figure}
\centerline{ \epsfxsize=500pt \epsfbox{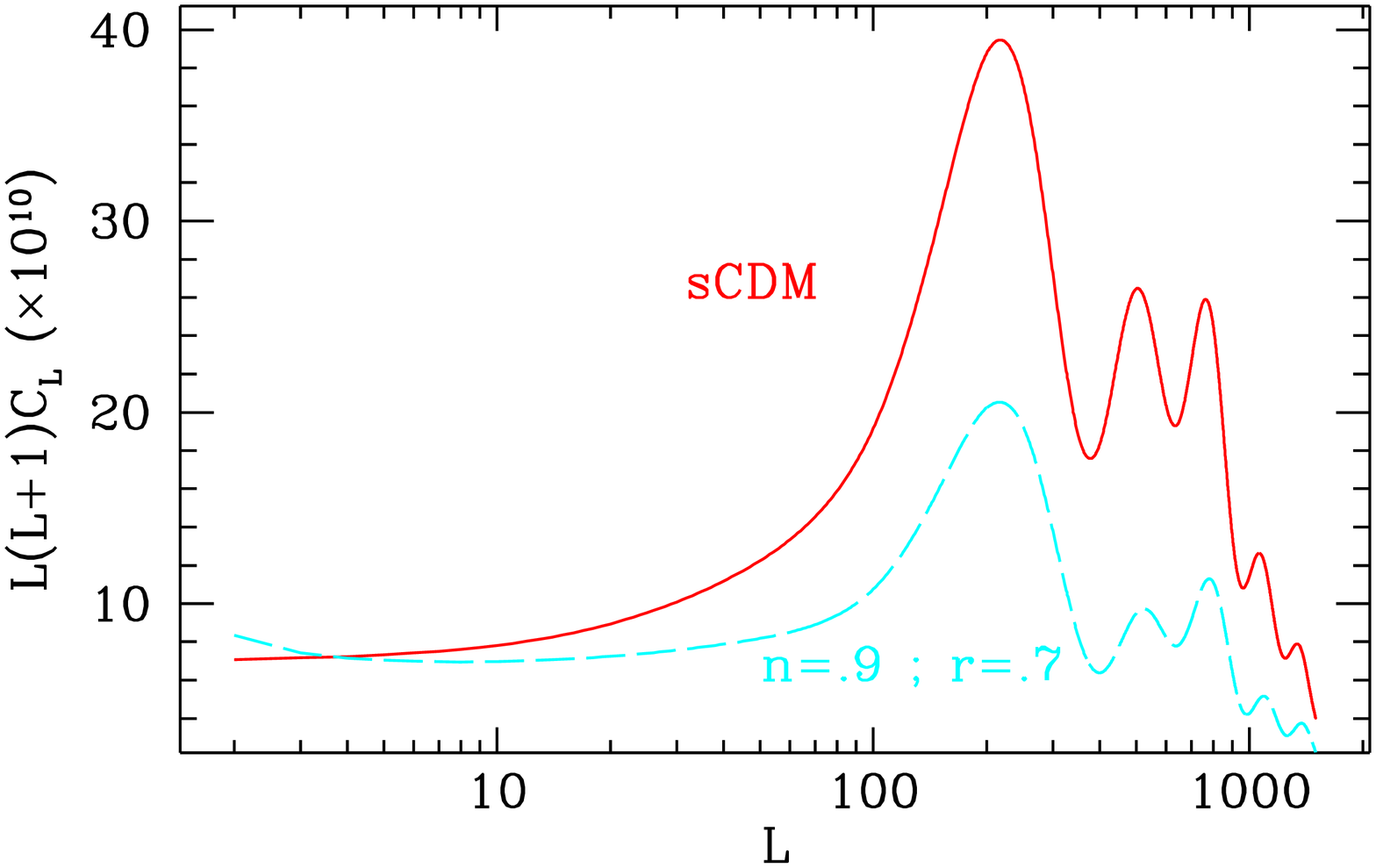} }
\footnotesize{\hspace*{0.2in} Fig.\ 1: The spectrum of anisotropies
for the two models discussed in the text. Both are normalized
at large scales to COBE. The model with $n=0.9$ is a much better
fit to large scale structure data.}
\end{figure}
%%%%%%%%%%%%%%%%%%%%%%%%%%%%%%%%%%%%%%%%%%%%
%%%%%%%%%%%%%%%%%%%%%%%%%%%%%%%%%%%%%%%%%%%%
%%%%%%%%%%%%%%%%%%%%%%%%%%%%%%%%%%%%%%%%%%%%

The last piece of information we need to compute the derivative matrix
in \eqr{EXCHI}\ is the uncertainty expected in the $C_l$'s. The
relevant experimental parameters are: the beam width, $\sigma_{\rm
beam}$; the expected noise per pixel, $\sigma_{\rm pixel}$; the area
per pixel, $\Omega_{\rm pixel}$; and the fraction of the sky covered.
Once these are known, it is very useful to employ a formula derived by
Knox \re{KNOX}, who showed that for an all-sky map,
\begin{equation}
{ \Delta C_l \over C_l} = \sqrt{2\over 2l+1} \left( 1 +
         {\sigma_{\rm pixel}^2 \Omega_{\rm pixel}\over C_l}
 \exp\{ l^2 \sigma_{\rm beam}^2 \}
        \right) \ .
\label{deltacl}
\end{equation}
The first term here is the inevitable consequence of the fact that we
have only $2l+1$ pieces of information at each $l$ (cosmic
variance). We will consider the MAP and PLANCK satellites. For MAP, we
assume $\sigma_{\rm beam} = 0.425 \times 0.3^\circ$ and $\sigma_{\rm
pixel}^2 \Omega_{\rm pixel} = (35\mu {\rm K})^2 (0.3^\circ)^2$. For
PLANCK, we take $\sigma_{\rm beam} = 0.425 \times 0.17^\circ$ and
$\sigma_{\rm pixel}^2 \Omega_{\rm pixel} = (3\mu {\rm K})^2
(0.167^\circ)^2$.

The results are shown in Fig.\ 2. The ellipses delineate $95\%$
confidence limits in $n$ and $r$ for the sCDM and LSS examples.  In
the sCDM case, we have imposed the (physical) restriction that
$r>0$. Also shown in Fig.\ 2 are the predictions from
the various models discussed in Section III.  By inverting
\eqrr{scalarspectralindex}{rofepsilon}, we can plot the same ellipses
in the $\eta-\epsilon$ plane. These are shown in Fig.\ 3.  The
superposition of the ellipses on top of the model predictions makes
clear that CMB observations will be able to discriminate amongst
inflationary models.

%%%%%%%%%%%%%%%%%%%%%%%%%%%%%%%%%%%%%%%%%%%%
%%%%%%%%%%%  FIGURE  2 %%%%%%%%%%%%%%%%%%%%%%%%%%%
%%%%%%%%%%%%%%%%%%%%%%%%%%%%%%%%%%%%%%%%%%%% 
\begin{figure}[p]
\centerline{ \epsfxsize=375pt \epsfbox{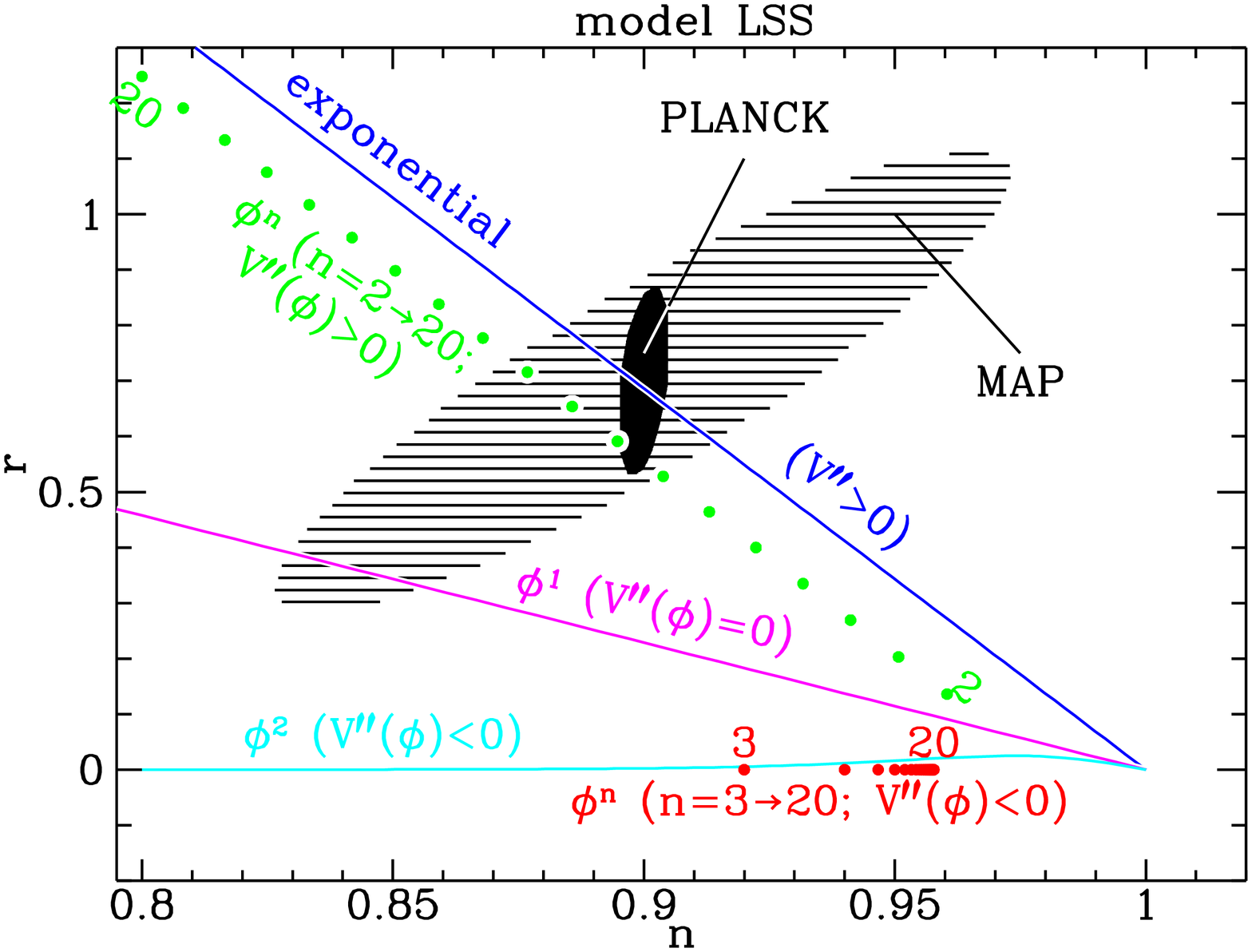} } \centerline{
\epsfxsize=375pt \epsfbox{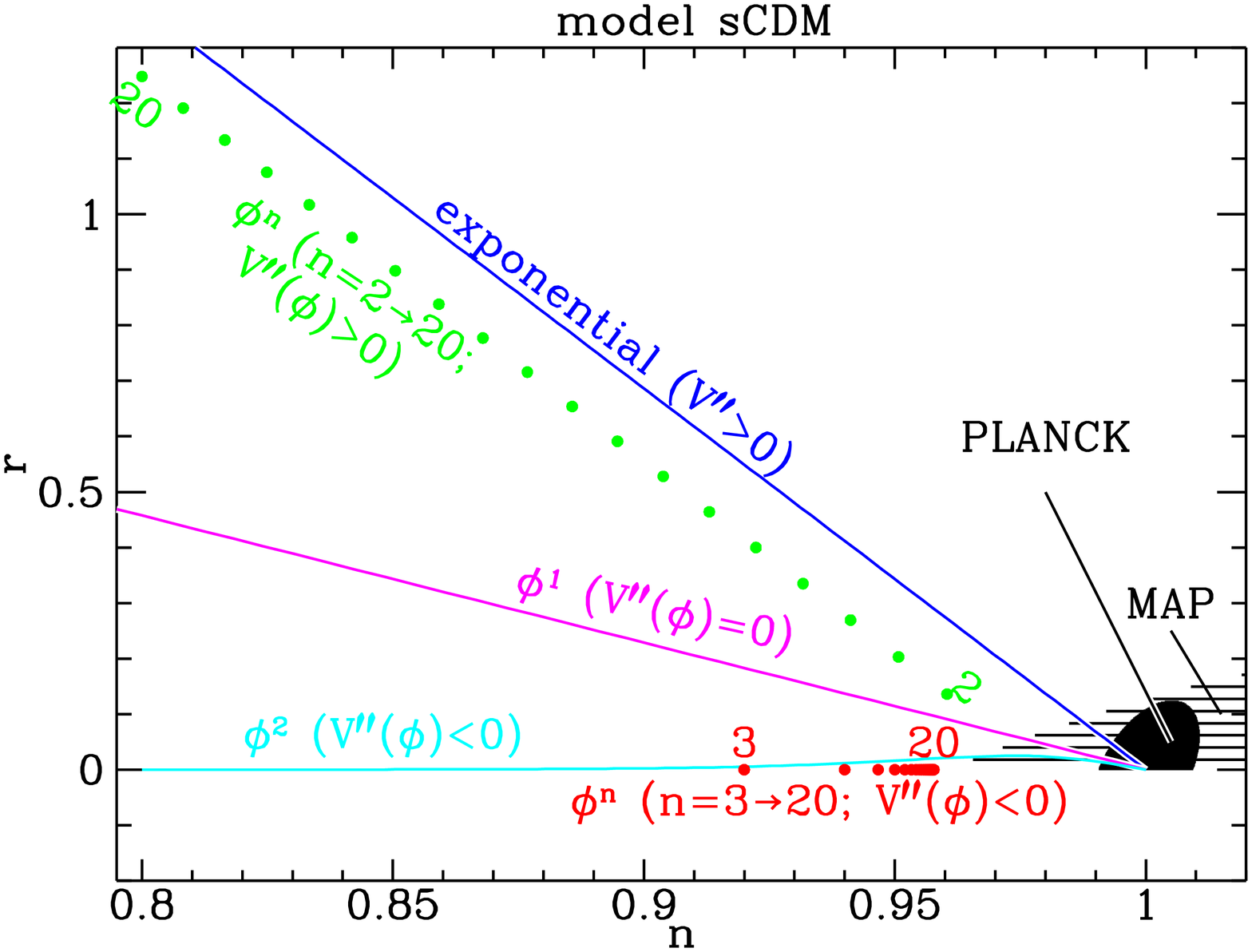} } 
\footnotesize{\hspace*{0.2in} Fig.\ 2: Predictions for a variety of
inflationary models in the $n-r$ plane superimposed on the expected
(95\% C.L.) region allowed by the two CMB satellites. The two panels
correspond to two different values of the true parameters: the upper
figure is the LSS model while the lower one is the sCDM model.  The
line labelled $\phi^1$ delineates two classes of models: Large (small)
field models lie above (below) the line.}
\end{figure}
%%%%%%%%%%%%%%%%%%%%%%%%%%%%%%%%%%%%%%%%%%%%
%%%%%%%%%%%%%%%%%%%%%%%%%%%%%%%%%%%%%%%%%%%%
%%%%%%%%%%%%%%%%%%%%%%%%%%%%%%%%%%%%%%%%%%%%

%%%%%%%%%%%%%%%%%%%%%%%%%%%%%%%%%%%%%%%%%%%%
%%%%%%%%%%%  FIGURE  3 %%%%%%%%%%%%%%%%%%%%%%%%%%%
%%%%%%%%%%%%%%%%%%%%%%%%%%%%%%%%%%%%%%%%%%%%
\begin{figure}[p]
\centerline{ \epsfxsize=375pt \epsfbox{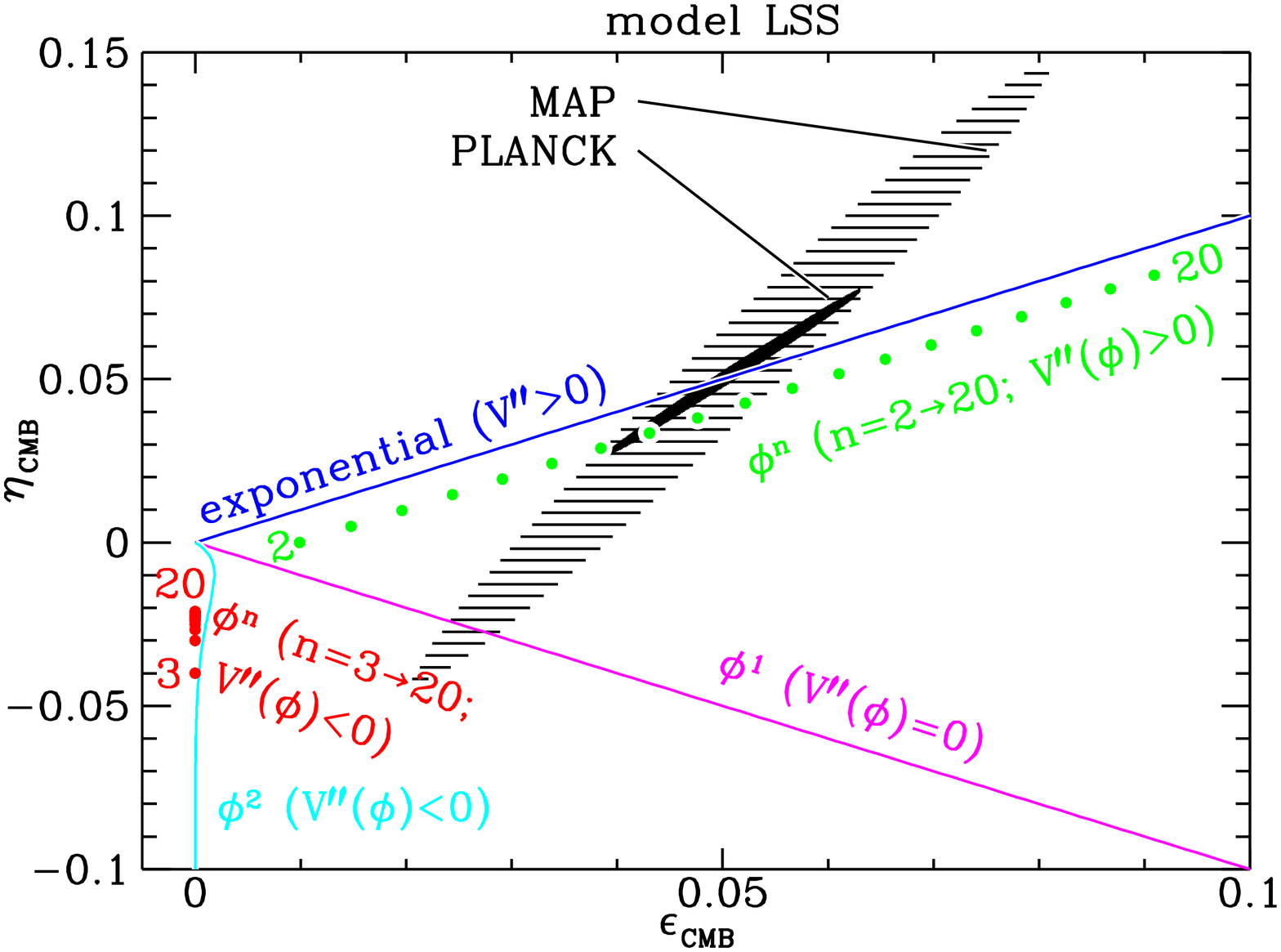} }
\centerline{ \epsfxsize=375pt \epsfbox{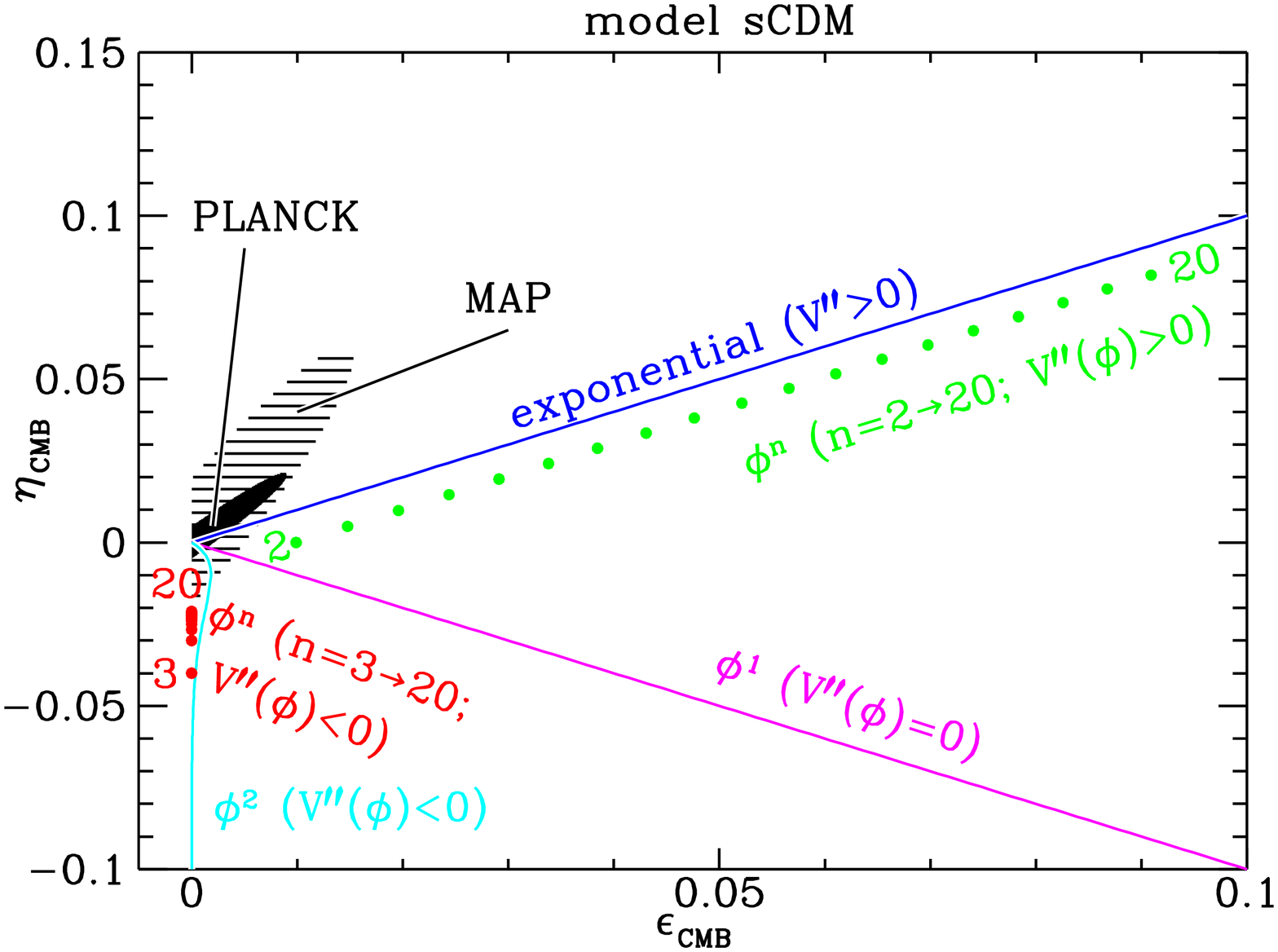} }
\footnotesize{\hspace*{0.2in} Fig.\ 3:  Same as Fig.\ 2, but now
the observational constraints have been mapped directly
onto the $\epsilon-\eta$ plane.}
\end{figure}
%%%%%%%%%%%%%%%%%%%%%%%%%%%%%%%%%%%%%%%%%%%%
%%%%%%%%%%%%%%%%%%%%%%%%%%%%%%%%%%%%%%%%%%%%
%%%%%%%%%%%%%%%%%%%%%%%%%%%%%%%%%%%%%%%%%%%%

\newpage

%%%%%%%%%%%%%%%%%%%%%%%%%%%%%%%%%%%%%%%%%%%%
\vspace{48pt}
\thesection{\centerline{\large \bf V. CONCLUSIONS}}
\setcounter{section}{5}
\setcounter{equation}{0}
\vspace{18pt}
%%%%%%%%%%%%%%%%%%%%%%%%%%%%%%%%%%%%%%%%%%%%
%%%%%%%%%%%%%%%%%%%%%%%%%%%%%%%%%%%%%%%%%%%%
%%%%%%%%%%%%%%%%%%%%%%%%%%%%%%%%%%%%%%%%%%%%

Different inflationary models make different predictions for the
spectrum of scalar and tensor perturbations.  While very different models
might lead to indistinguishable scalar spectra, it has been realized
for some time that the tensor spectrum, used in conjunction with the
scalar spectrum, can differentiate between models \re{ckll}.  Here we
have demonstrated how the effect of scalar and tensor combinations on
CMB fluctuations can be used as a discriminant in testing inflation
models.

Most inflationary models have an adjustable parameter that can be
tuned to give the correct normalization of the scalar perturbations ($\qrms$ in
the language used to study CMB fluctuations).  A simple example of
such a parameter is the coupling constant $\lambda$ in the chaotic
inflation model with potential $V(\phi) = \lambda \phi^4$.  However,
in this paper we have shown that even with the freedom of an
adjustable parameter it is possible that observations of the cosmic
microwave background can distinguish among different inflation models.
Therefore, we can hope in the next decade to see a real confrontation
between inflation models and CMB observations.

While the type of analysis we propose can never ``prove'' that any
particular model is correct, it might do much more than simply
eliminate models.  It is possible that an analysis like the one we
present here might be able to give some guidance in model building.
One way of dividing inflationary models is to classify them as either
``small-field'' models, ``large-field'' models, or ``hybrid''
models.\footnote{A more exact division would be according to the
second derivative of the potential near $\phi_{\rm CMB}.$} Different
versions of the three types of models predict qualitatively different
scalar and tensor spectra, so it should be particularly easy to tell
them apart once the data is available.

Although we have only studied simple examples of models, we can
speculate that small-field, large-field, and hybrid models will
populate different regions of the $n$--$r$ plane as illustrated in
Fig.\ 4.  Certainly a scalar spectral index larger than unity would
suggest some form of hybrid model.  A scalar index smaller than one in
combination with negligible tensor contribution (small $r$) would
suggest a small-field model, while scalar index less than unity with
considerable tensor contribution would point toward large-field
models.

%\newpage

%%%%%%%%%%%%%%%%%%%%%%%%%%%%%%%%%%%%%%%%%%%%
%%%%%%%%%%%  FIGURE  4 %%%%%%%%%%%%%%%%%%%%%%%%%%%
%%%%%%%%%%%%%%%%%%%%%%%%%%%%%%%%%%%%%%%%%%%%
\begin{figure}
\centerline{ \epsfxsize=450pt \epsfbox{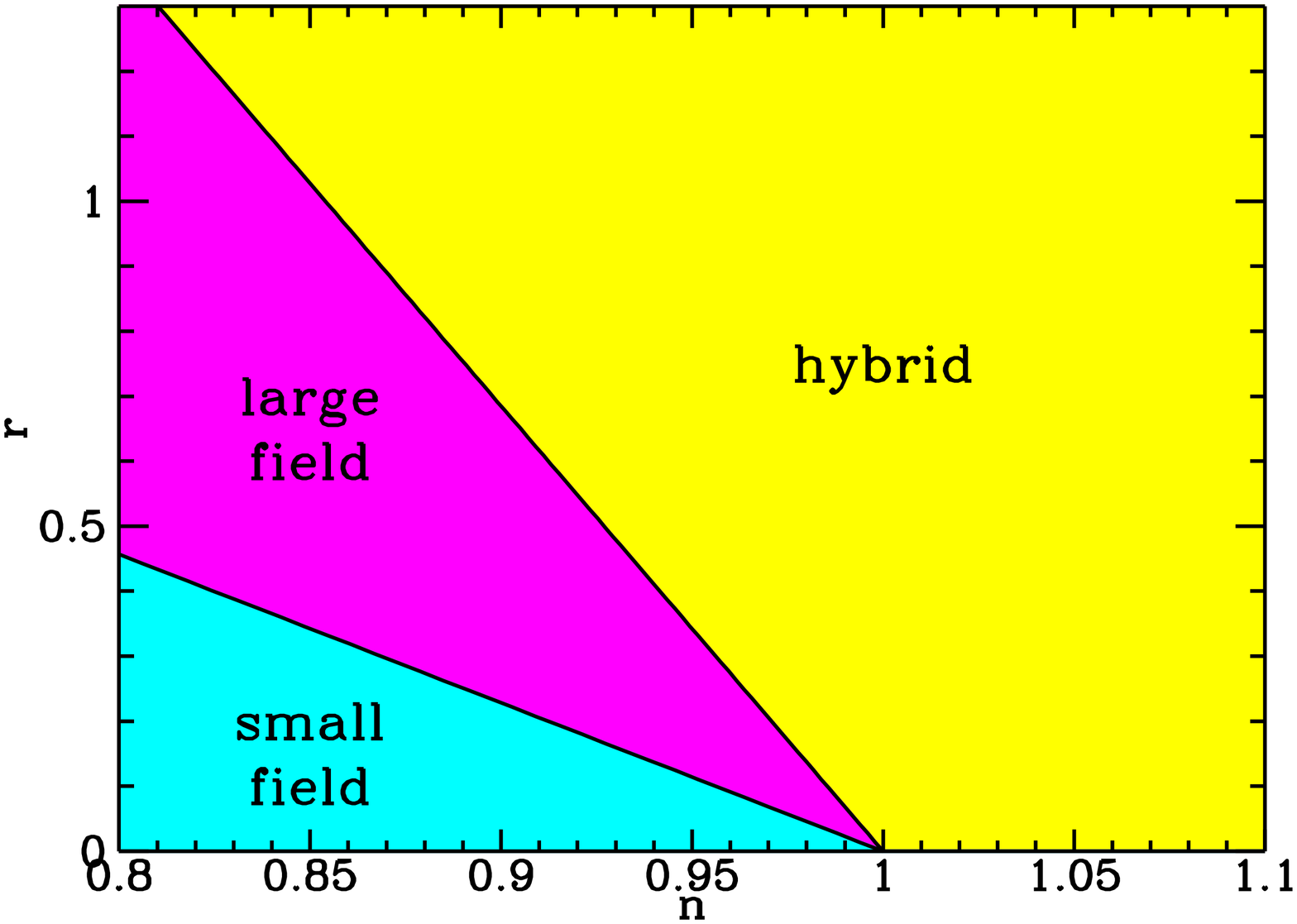} }
\footnotesize{\hspace*{0.2in} Fig.\ 4:  Regions in the $n$--$r$ plane 
populated by the three types of models considered in this paper.}
\end{figure}
%%%%%%%%%%%%%%%%%%%%%%%%%%%%%%%%%%%%%%%%%%%%
%%%%%%%%%%%%%%%%%%%%%%%%%%%%%%%%%%%%%%%%%%%%
%%%%%%%%%%%%%%%%%%%%%%%%%%%%%%%%%%%%%%%%%%%%

An interesting question we do not address here is whether
single-field, slow-roll models populate the entire $n$--$r$ plane.

%%%%%%%%%%%%%%%%%%%%%%%%%%%%%%%%%%%%%%%%%%%%
%%%%%%%%%%%%%%%   ACKNOWLEDGEMENTS  %%%%%%%%%%%%%%%
%%%%%%%%%%%%%%%%%%%%%%%%%%%%%%%%%%%%%%%%%%%%
\vspace{36pt}

We thank Uros Seljak and Matias Zaldariagga for use
of their CMBFAST code \re{SEL}.
This work was supported in part by DOE and NASA grant NAG5--2788 at Fermilab.
%%%%%%%%%%%%%%%%%%%%%%%%%%%%%%%%%%%%%%%%%%%%
%%%%%%%%%%%%%%%%%%%%%%%%%%%%%%%%%%%%%%%%%%%%
%%%%%%%%%%%%%%%%%%%%%%%%%%%%%%%%%%%%%%%%%%%%

\newpage 
%%%%%%%%%%%%%%%%%%%%%%%%%%%%%%%%%%%%%%%%%%%%
%%%%%%%%%%%%%%%%  REFERENCES    %%%%%%%%%%%%%%%%%%%
%%%%%%%%%%%%%%%%%%%%%%%%%%%%%%%%%%%%%%%%%%%%

\begin{picture}(400,50)(0,0)
\put (50,0){\line(350,0){300}}
\end{picture}

\vspace{0.25in}

\def\labelenumi{[\theenumi]}
\frenchspacing
\def\prl{{{\em Phys. Rev. Lett.\ }}}
\def\prd{{{\em Phys. Rev. D\ }}}
\def\pl{{{\em Phys. Lett.\ }}}

\begin{enumerate}

\item\label{MAP} http://map.gsfc.nasa.gov/

\item\label{CS} 
http://astro.estec.esa.nl/SA-general/Projects/Cobras/cobras.html

\item\label{CT} R. G. Crittenden and N. G. Turok, {\em
Phys. Rev. Lett.} {\bf 75}, 2642 (1995).

\item\label{ANDY} A. Albrecht, D. Coulson, P. Ferreira, and
J. Magueijo, {\em Phys. Rev. Lett.} {\bf 76}, 1413 (1996).

\item\label{HW} W. Hu and M. White, {\em Phys. Rev. Lett.}  {\bf 77},
1687 (1996)

\item\label{TUROK} N. G. Turok, astro-ph/9607109 (1996).

\item\label{KNOX} L. Knox, {\em Phys. Rev. D} {\bf 52}, 4307 (1995).

\item\label{KTURN} L. Knox and M. S. Turner, {\em Phys. Rev. Lett.}
{\bf 73}, 3347 (1994).

\item\label{COSCON} J. R. Bond, R. Crittenden, R. L. Davis, G.
       Efstathiou and P. J. Steinhardt, {\em Phys. Rev.  Lett.}  {\bf
       72}, 13 (1994).

\item\label{JUNG} G. Jungman, M. Kamionkowski, A.
       Kosowsky , D. N. Spergel {\em Phys. Rev. D} {\bf 54}, 1332 (1996).

\item\label{ME} S. Dodelson, E. I. Gates, and A. S. Stebbins, 
{\em Astrophys. J.} {\bf 467}, 10 (1996).

\item\label{transfer} See, e.g. G. Efstathiou in {\em Physics of the
Early Universe}, edited by J. A. Peacock, A. F. Heavens, and
A. T. Davies (Adam Higler, Bristol, 1990).

\item\label{llkcba97} J. E. Lidsey, A. R. Liddle, E. W. Kolb, E. J. Copeland, 
T. Barriero, and M. Abney, {\em Rev. Mod. Phys.}, (1997).

\item\label{gs88} L. P.  Grishchuk and Yu. V. Sidorav,  
	in {\it Fourth Seminar on 
	Quantum Gravity}, eds M. A. Markov, V. A. Berezin and V. P. 
	Frolov (World Scientific, Singapore, 1988)

\item\label{m90} A. G. Muslimov, {\em Class. Quant. Grav.} {\bf 7}, 231 
(1990).

\item\label{sb90}  D. S. Salopek and J. R. Bond, {\em Phys. Rev. D} {\bf 42}, 
3936 (1990).

\item\label{kv94} E. W. Kolb and S. L. Vadas, {\em Phys. Rev. D}, {\bf 50},
2479 (1994).

\item\label{twl93} M. S. Turner, M. White and J. E. Lidsey, {\em Phys. Rev. 
D} {\bf 48}, 4613 (1993).

\item\label{al91} A. D. Linde, {\em Phys. Lett.} {\bf 259B}, 38 (1991).

\item\label{al94} A. D. Linde, {\em Phys. Rev. D.} {\bf 49}, 748 (1994).

\item\label{cllsw94} E. J. Copeland, A. R. Liddle, D. H. Lyth, E. D. Stewart 
and D. Wands, {\em Phys. Rev. D} {\bf 49} 6410 (1994).

\item\label{al82} A. D. Linde, {\em Phys. Lett.} {\bf 108B}, 389 (1982).

\item\label{as82} A. Albrecht and P. J. Steinhardt, {\em
Phys. Rev. Lett.} {\bf 48}, 1220 (1982).

\item\label{km96} W. H. Kinney and K. T. Mahanthappa, {\em Phys. Rev. D}
{\bf 53}, 5455 (1996).

\item\label{NATURALINFLATION} K. Freese, J. Frieman, and A. Olinto,
{\em Phys. Rev. Lett.} {\bf 65}, 3233 (1990).

\item\label{PRESS} 
 W. H. Press,  S. A. Teukolsky,  W. T. Vetterling, and  B. P. Flannery, 
{\em Numerical Recipes}, (Cambridge: Cambridge University Press, 1992).

\item\label{EQUI} M. White, D. Scott, J. Silk, M. Davis,
{\em Mon. Not. Roy. Astron. Soc.} {\bf 276}, L69 (1995).

\item\label{ckll} E. Copeland, E. W. Kolb, A. R. Liddle, and
J. E. Lidsey, {\it Phys. Rev. Lett.} {\bf 71}, 219 (1993) and {\it
Phys. Rev. D} {\bf 48}, 2529 (1993).

\item\label{SEL} U. Seljak and M. Zaldariagga,
{\em Astrophys. J.} {\bf 469}, 7 (1996).

\end{enumerate}

\end{document}